\newcommand{\be}{\begin{equation}}\newcommand{\ee}{\end{equation}}
\newcommand{\bea}{\begin{eqnarray}}\newcommand{\eea}{\end{eqnarray}}
\newcommand{\nn}{\nonumber}\newcommand{\p}[1]{(\ref{#1})}
\newcommand{\lb}{\label}

\newcommand{\cA}{{\cal A}}
\newcommand{\bcA}{\bar{\cal A}}

\newcommand\T{\mbox{Tr}\;}

\newcommand\s{\scriptscriptstyle}
\newcommand\q{\quad}
\newcommand\qq{\qquad}

\newcommand{\pp}{{=\!\!\!|}}
\newcommand{\xp}{x^\pp}
\newcommand{\xm}{x^=}

 \newcommand{\Pp}{\partial_\pp}
\newcommand{\Pm}{\partial_=}

\newcommand{\sY}{{\s Y}}
 \newcommand{\PY}{\partial_\sY}
 \newcommand{\bPY}{\bar\partial_\sY}

\newcommand{\tpi}{\theta^+_i}

\newcommand{\tmi}{\theta^-_i}

\newcommand{\btpi}{\bar{\theta}^{i+}}

\newcommand{\btmi}{\bar{\theta}^{i-}}

\newcommand{\bppt}{\bar{\partial}_{2+}}

\newcommand{\bpmt}{\bar{\partial}_{2-}}
\newcommand{\bpmh}{\bar{\partial}_{3-}}
\newcommand{\bpph}{\bar{\partial}_{3+}}
\newcommand{\ppo}{\partial^1_+}
\newcommand{\ppt}{\partial^2_+}
\newcommand{\pmo}{\partial^1_-}
\newcommand{\pmt}{\partial^2_-}

\newcommand{\Dpk}{D_+^k}
\newcommand{\Dpl}{D_+^l}

\newcommand{\Dmk}{D_-^k}
\newcommand{\Dml}{D_-^l}

\newcommand{\bDpk}{\bar{D}_{k+}}
\newcommand{\bDpl}{\bar{D}_{l+}}

\newcommand{\bDmk}{\bar{D}_{k-}}
\newcommand{\bDml}{\bar{D}_{l-}}

\newcommand{\tpo}{\theta^+_1}
\newcommand{\tpt}{\theta^+_2}
\newcommand{\tmo}{\theta^-_1}
\newcommand{\tmt}{\theta^-_2}
\newcommand{\tmh}{\theta^-_3}
\newcommand{\tph}{\theta^+_3}
\newcommand{\btpo}{\bar\theta^{1+}}
\newcommand{\btmo}{\bar\theta^{1-}}
\newcommand{\btpt}{\bar\theta^{2+}}
\newcommand{\btmt}{\bar\theta^{2-}}
\newcommand{\btph}{\bar\theta^{3+}}
\newcommand{\btmh}{\bar\theta^{3-}}
\newcommand{\Dpo}{D_+^1}
\newcommand{\Dmo}{D_-^1}
\newcommand{\Dpt}{D_+^2}

\newcommand{\bDph}{\bar{D}_{3+}}
\newcommand{\bDmh}{\bar{D}_{3-}}

\newcommand{\Dot}{D^1_2}
\newcommand{\Dto}{D^2_1}
\newcommand{\Doh}{D^1_3}

\newcommand{\Dth}{D^2_3}

\newcommand{\hDot}{\hat{D}^1_2}

\newcommand{\hDoh}{\hat{D}^1_3}

\newcommand{\hDth}{\hat{D}^2_3}

\documentstyle[epsfig,12pt]{article}

\begin{document}
\setlength{\baselineskip}{5mm}
\hfill hep-th/0010139
\vspace{1cm}

\noindent{\large\bf SOLVING N=3 SUPER-YANG-MILLS EQUATIONS\\ IN HARMONIC 
SUPERSPACE\footnote{Talk at the 23 International Colloquium 
``Group-Theoretical Methods in Physics'', Dubna,  2000}
}\vspace{4mm}

\noindent{\hspace{1cm}\bf B.M. ZUPNIK 
}\vspace{1mm}

\noindent{\small
 Bogoliubov Laboratory of Theoretical Physics, Joint Institute
for Nuclear Research, Dubna, Moscow Region, 141980, Russia; e-mail:
zupnik@thsun1.jinr.ru
}\vspace{4mm}

{\small
We analyze the superfield constraints of the $D{=}4,~N{=}3$ SYM-theory 
using light-cone gauge conditions. The $SU(3)/U(1)\times U(1)$ harmonic 
variables are interpreted as auxiliary spectral parameters, and the
transform to the  harmonic-superspace representation is considered. Our 
nilpotent gauge for the basic harmonic superfield simplifies the 
SYM-equations of motion. A partial Grassmann decomposition of these
equations yields the solvable linear system of iterative equations.
}

\section{Introduction}

It is well known that the geometric superspace constraints of the $D=4, N=3$
SYM-theory are equivalent to the equations of motion for the component
fields [1]. The special projections of these superfield constraints
can be interpreted as  conditions of zero curvatures assosiated with
the Grassmann  covariant derivatives [2,3]. The possible connection of these 
zero-curvature conditions with integrability or solvability of the 
super-Yang-Mills (SYM) theories with 16 or 12 supercharges has been discussed 
more than 20 years in the framework of different superfield approaches (see, 
e.g. [2-6]). Here we shall use the harmonic-superspace approach [3] to 
analyze the solutions of the $N=3$ SYM-equations.

The harmonic-superspace (HSS) method has been introduced first for the 
solution of the $D=4,~N=2$ off-shell superfield constraints [7]. Harmonic 
variables are analogous to  twistor or  auxiliary spectral variables used in  
integrable models. Harmonic and twistor methods give the explicit 
constructions of the general solutions to zero-curvature conditions in terms 
of independent functions on special {\it analytic} (super)spaces satisfying 
the generalized Cauchy-Riemann  analyticity conditions. For instance, the 
off-shell $N=2$ superfields satisfy the conditions of the Grassmann (G-) 
analyticity.

In the standard harmonic formulation of the  $D=4, N=2$ SYM-theory,
the basic harmonic connection is G-analytic [7], and the 2-nd one ( via the 
zero-curvature condition) appears to be a nonlinear function of
the basic connection. The $N=2$ equation of motion is linearly dependent on 
the 2-nd harmonic connection, but it is the nonlinear equation for the basic 
connection [8]. It has been shown in Ref.[5] that one can alternatively 
choose the 2-nd harmonic connection as a  basic superfield, so that
the dynamical G-analyticity condition (or the Grassmann-harmonic 
zero-curvature condition) for the first connection becomes a new equation of 
motion. We shall use below a similar change of the basic HSS 
variables for the $N=3$ SYM-theory.

In the HSS-approach to the $D=4,\;N=3$ supersymmetry, the $SU(3)/U(1)\times 
U(1)$ harmonics have been used for the covariant reduction of the spinor 
coordinates and derivatives and for the off-shell description of the  
SYM-theory in terms of the corresponding G-analytic superfields [3].
The $N=3$ SYM-equations in the ordinary superspace  have been
transformed to the zero-curvature equations for the harmonic
gauge connections, however, nobody tried earlier to obtain the
general solution of these equations.

The alternative  $SL(2,C)$-harmonic interpretation of the $N=3$
SYM-equations and the corresponding harmonic zero-curvature equation for 
two G-analytic connections has been considered in Ref.[6]. We shall not 
discuss in this paper the problem of solving the equations of the  
$SL(2,C)$-harmonic approach.

An analogy with the self-dual SYM-solution [9] has been used
recently for the analysis of  the $10D$ SYM-solutions [10].  
These authors have considered the very useful light-cone gauge conditions 
which break the space-time symmetry of the SYM-equations. We shall
consider the analogous light-cone gauge conditions which preserve the 
$SU(3)$-invariance and allow us to use the harmonic approach.

It should be underlined that  the $N=3$ harmonic equations of motion
contain three analytic connections, so they are more complicated than the 
$SU(2)$-harmonic  equations using the single analytic connection.
If one does not treat the $SU(3)$-harmonic connections as the basic
variables, then the harmonic zero-curvature equations can be readily
solved in terms of the non-analytic superfield  matrix $v$ ( bridge).
In this approach, the   G-analyticity conditions  for the composed harmonic
connections can be considered as the dynamical equations
for this independent superfield $v$.

 It will be shown that  the light-cone gauge conditions for the
initial superfield connections correspond to the light-cone
analyticity of the bridge matrix. We choose the nilpotent gauge for $v$
and construct the nilpotent analytic harmonic connections in this
representation. We demonstrate that the 1-st order harmonic bridge equations 
with the nilpotent connections produce also linear 2-nd order differential 
constraints for the basic matrices.

This talk is based on our work [11].

\section{\lb{B}Reduced-symmetry representation of $D=4,~N=3$ SYM
 constraints}

The covariant coordinates  of the $D=4,~N=3$ superspace are
\be
z^M=(x^{\alpha\dot\alpha} ,\theta^\alpha_i ,\bar\theta^{i\dot\alpha} )~,
\lb{A2b}
\ee
where $\alpha,~\dot\alpha$ are the $SL(2,C)$ indices
and $i=1, 2, 3$  are indices of the fundamental
representations of the  group $SU(3)$.

We shall study solutions of the  SYM-equations using the non-covariant
representation  and the new  notation for
 these coordinates
\bea
&&\xp\equiv x^{1\dot{1}} =t+x^3~,\q \xm\equiv x^{2\dot{2}} =t-x^3~,\q
y\equiv x^{1\dot{2}}=x^1+ix^2~,\q\bar{y}\equiv x^{2\dot{1}} =x^1-ix^2~,
\nn\\
&&(\tpi,~\tmi)\equiv\theta^\alpha_i~,\q (\btpi,~\btmi)\equiv
 \bar\theta^{i\dot\alpha}
~.\lb{A2}
\eea
based on  the reduced symmetry $SO(1,1)\times SU(3)$.

 The $SO(1,1)$ weights (helicities) of these coordinates are
\be
w(\xp)=2~,\q w(\xm)=-2~,\q w(y)=w(\bar{y})=0~,\q w(\theta^\pm_i)=
w(\bar\theta^{i\pm})=\pm1~.
\lb{A3}
\ee

 The reduced representation of the algebra of $D=4,~N=3$ spinor
derivatives is
\bea
&&\{\Dpk ,\Dpl\}=0~,\qq\{\bDpk ,\bDpl\}=0~,\q
 \{\Dpk ,\bDpl\}=2i\delta^k_l\Pp~,\nn\\
&&\{\Dmk ,\Dml\}=0~,\qq\{\bDmk ,\bDml\}=0~,\q
 \{\Dmk ,\bDml\}=2i\delta^k_l\Pm~,\lb{A6}\\
&&
 \{\Dpk ,\bDml\}=2i\delta^k_l\partial_y~,\q
\{\Dmk ,\bDpl\}=2i\delta^k_l\bar\partial_y~,
\q\{\Dpk ,\Dml\}=\{\bDpk ,\bDml\}=0~.\nn
\eea
The last two relations can contain, in principle, six central charges,
however, we shall consider the basic superspace  without central charges.
The general $N=3$ superspace has the odd dimension (6,6) in our
notation with the reduced $SO(1,1)$ symmetry.

Let us consider the $(4|6,6)$-dimensional superspace gauge connections
$A(z)$ and the corresponding covariant derivatives $\nabla$
\bea
&&\nabla^i_\pm=D^i_\pm + A^i_\pm~,\qq\bar{\nabla}_{i\pm}=\bar{D}_{i\pm} +
\bar{A}_{i\pm}~,\lb{A7}\\
&&\nabla_\pp =\Pp + A_\pp ~,\qq\nabla_= =\Pm + A_= ~,\q
\nabla_y=\partial_y+A_y~,\q \bar\nabla_y=\bar\partial_y+\bar{A}_y~.\nn
\eea

 The $D=4,~N=3$ SYM-constraints [1]
have the following reduced-superspace form:
\bea
&&\{\nabla^k_+,\nabla^l_+\}=0~,\q \{\bar{\nabla}_{k+},\bar{\nabla}_{l+}\}
=0~,
\q \{\nabla^k_+,\bar{\nabla}_{l+}\}=2i\delta^k_l\nabla_\pp~,\lb{A8}\\
&&\{\nabla^k_+,\nabla^l_-\}=\bar{W}^{kl}
~,\q\{\nabla^k_+,\bar{\nabla}_{l-}\}=2i\delta^k_l \nabla_y~,\lb{A9}\\
&&\{\nabla^k_-,\bar{\nabla}_{l+}\}=2i\delta^k_l\bar{\nabla}_y~,\q
\{\bar{\nabla}_{k+},\bar{\nabla}_{l-}\}=W_{kl}~,\lb{A10}\\
&&\{\nabla^k_-,\nabla^l_-\}=0~,\q \{\bar{\nabla}_{k-},\bar{\nabla}_{l-}\}
=0~,
\q \{\nabla^k_-,\bar{\nabla}_{l-}\}=2i\delta^k_l\nabla_=~,\lb{A11}
\eea
where $W_{kl}$ and $\bar{W}^{kl}$ are the gauge-covariant
superfields constructed from the gauge connections. This reduced form
of the 4D constraints is convenient for the study of dimensionally
reduced SYM-subsystems in $D<4$.

Let us analyze first Eqs.\p{A8} combined with the relations
\be
[\nabla^k_+,\nabla_\pp]=[\bar\nabla_{k+},\nabla_\pp]=0~.
\ee

The simplest light-cone gauge condition is
\be
A^k_+=0~,\q \bar{A}_{k+}=0~,\q A_\pp =0~.\lb{A14}
\ee
The analogous gauge conditions excluding the part of connections has
been considered in Ref.[9] for the self-dual $4D$ SYM-theory and
in Ref.[10] for the 10D SYM equations.

Stress that the light-cone gauge preserves the $SU(3)$-symmetry of the
$N=3$ superfield constraints, so we can use the harmonic-superspace
method for  non-trivial equations (\ref{A9}-\ref{A11}) in this gauge.

\section{\lb{D}Harmonic-superspace equations for the nilpotent
 bridge matrix}

The $SU(3)/U(1)\times U(1)$ harmonic superspace has been introduced
in Ref.[3] for the off-shell description of the $4D\; N=3$
SYM-theory. The dynamical equations in this approach have been transformed
into the set of pure harmonic equations for the G-analytic superfield
prepotentials. It should be stressed that nobody tried earlier to analyze
in details the solutions of these harmonic equations.

Now we shall  study the reduced-symmetry version
of the $SU(3)/U(1)\times U(1)$ harmonic superspace which allows us
to consider the non-covariant gauges and the dimensional reduction.

The $SU(3)/U(1)\times U(1)$ harmonics [3] parameterize
the corresponding coset space. They form an $SU(3)$ matrix $u^I_i$ and
are defined modulo $U(1)\times U(1)$ transformations
\be
u^1_i=u^{(1,0)}_i\;,\q u^2_i=u^{(-1,1)}_i\;,\q
u^3_i=u^{(0,-1)}_i\;,\lb{A15}
\ee
where the lower index $i$ describes the triplet representation of $SU(3)$,
and the upper indices 1, 2 and 3 correspond to different combinations
of charges. The
complex conjugated harmonics have opposite $U(1)$ charges and reverse
positions of indices
\be
u^i_1=u^{i(-1,0)}~,\q u^i_2=u^{i(1,-1)}\;,\q u^i_3=u^{i(0,1)}\;.
\lb{A16}
\ee

These harmonics satisfy the following relations:
\bea
&& u_i^I u^i_J=\delta^I_J\;,\q u^I_i u^k_I=\delta^k_i\;,\nn\\
&&\varepsilon^{ikl}u_i^1 u_k^2 u_l^3=1\;.\lb{A17}
\eea

The $SU(3)$-invariant harmonic derivatives act on the harmonics
\bea
&&\partial^I_J u^K_i =\delta^K_J u^I_i\;,\q \partial^I_J u^i_K=-\delta^I_K
u^i_J\;,\nn\\
&&[\partial^I_J,\partial^K_L]=\delta^K_J\partial^I_L-\delta^I_L\partial^K_J
\;.\lb{A18}
\eea

We shall use the special $SU(3)$-covariant conjugation
\be
\widetilde{u^1_i}=u^i_3\;,\q \widetilde{u^3_i}=u_1^i\;,\q
\widetilde{u^2_i}=-u_2^i\;.\lb{A19}
\ee

We can define the real analytic harmonic superspace $H(4,6|4,4)$
with 6 coset harmonic dimensions and the following left and right
coordinates:
\bea
&&\zeta=(X^\pp~,~X^=~,~Y~,~\bar{Y}~,~ \theta^\pm_2~,
~\theta^\pm_3~,~\bar\theta^{1\pm}~,~\bar\theta^{2\pm})~,\nn\\&&
X^\pp=\xp +i(\tph\btph -\tpo\btpo)\;,\qq X^= =\xm +
i(\tmh\btmh -\tmo\btmo)\;,\nn\\
&& Y=y+i(\tph\btmh -\tpo\btmo)\;,\qq\bar{Y}=\bar{y}
+i(\tmh\btph -\tmo\btpo)\;,\lb{A22}
\eea
where
$\theta^\pm_I=\theta^\pm_k u^k_I~,\q\bar\theta^{\pm I}=
\bar\theta^{\pm k}u_k^I$.

The  CR-structure in $H(4,6|4,4)$ involves the G-derivatives
\be
D^1_\pm,\;\bar{D}_{3\pm}~,
\lb{A23}
\ee
which commute with the harmonic derivatives $\Dot,\;\Dth$ and $\Doh$.

These derivatives have the following explicit form in the
analytic coordinates:
\bea
&&D^{(1,0)}_\pm\equiv D^1_\pm=\partial^1_\pm\equiv
\partial/\partial\theta^\pm_1\;\q
\bar{D}^{(0,1)}_\pm\equiv \bar{D}_{3\pm}=\partial_{3\pm}\equiv
\partial/\partial \bar\theta^{3\pm}\;,\lb{A24}\\
&&D^{(2,-1)}\equiv \Dot =\partial^1_2
+i\tpt\btpo\Pp+i\tpt\btmo\PY+i\tmt\btpo\bPY
+i\tmt\btmo\Pm\nn\\
&&-\tpt\ppo-\tmt\pmo+\btpo\bppt+\btmo\bpmt
~,\nn\\
&&D^{(-1,2)}\equiv\Dth =\partial^2_3
+i\tph\btpt\Pp+i\tph\btmt\PY+i\tmh\btpt\bPY
+i\tmh\btmt\Pm\nn\\
&&-\tph\ppt-\tmh\pmt+\btpt\bpph+\btmt\bpmh\;,\lb{A25}\\
&&D^{(1,1)}\equiv\Doh =\partial^1_3
+2i\tph\btpo\Pp+2i\tph\btmo\PY+2i\tmh\btpo\bPY
+2i\tmh\btmo\Pm\nn\\
&&-\tph\ppo-\tmh\pmo+\btpo\bpph+\btmo\bpmh\;,\nn
\eea
where $\Pp =\partial/\partial X^\pp,~\Pm =\partial/
\partial X^=,~\PY=\partial/\partial Y$ and $\bPY=
\partial/\partial\bar{Y}$.

It is crucial that we start from  the light-cone  gauge conditions
\p{A14} for the $N=3$ SYM-connections which break  $SL(2,C)$ but preserve
the $SU(3)$-invariance. Consider the harmonic transform
 of the  covariant Grassmann derivatives in this gauge using the projections
on the $SU(3)$-harmonics
\bea
&&\nabla^I_+\equiv u_i^I D^i_+=D^I_+\;,\q
\bar\nabla_{I+}\equiv u^i_I\bar{D}_{i+}=\bar{D}_{I+}\;,\q
\{D^I_+,\bar{D}_{K+}\}=2i\delta^I_K\Pp\;,\lb{D1b}\\
&&\nabla^I_-\equiv u_i^I\nabla^i_-=D^I_- +\cA^I_-\;,\q
\bar\nabla_{I-}\equiv u^i_I\nabla_{i-}=\bar{D}_{I-} +\bcA_{I-}\;,
\lb{D1}
\eea
where the harmonized Grassmann connections $\cA^I_-$ and $\bcA_{I-}$
are defined.

The $SU(3)$-harmonic projections of the superfield constraints
(\ref{A9}-\ref{A11}) can be derived from the basic set of the
$N=3$ G-integrability conditions for two components
of the harmonized connection:
\bea
&&\Dpo\cA^1_-=\bDph\cA^1_-=\Dpo\bcA_{3-}=\bDph\bcA_{3-}=0\;\lb{D2}\\
&&\Dmo\cA^1_- +(\cA^1_-)^2=0\;,\q \bDmh\bcA_{3-}+(\bcA_{3-})^2=0\;,\nn\\
&&\Dmo\bcA_{3-}+\bDmh\cA^1_- +\{\cA^1_-,\bcA_{3-}\}=0\;.\lb{D3}
    \eea
All projections of the SYM-equations can be obtained by the action
of  harmonic  derivatives $D^I_K$ on these basic conditions.

These Grassmann zero-curvature equations have the very simple general
solution
\be
\cA^1_-(v)=e^{-v}\Dmo e^v\;,\q \bcA_{3-}(v)=e^{-v}\bDmh e^v\;,\lb{D4}
\ee
where {\it the bridge } $v$ is the  matrix in the Lie algebra 
of the gauge group. This superfield matrix satisfies the additional  
constraint
\be
(\Dpo,\;\bDph) v=0\;,\lb{D5}
\ee
which is compatible with the light-cone representation \p{D1b}. Thus,
$v$ does not depend on the Grassmann coordinates $\tpo$ and $\btph$.

Consider the gauge transformations of the bridge
\be
e^v\;\Rightarrow\;e^\lambda e^v e^{\tau_r}\;,\lb{D11}
\ee
where $\lambda\in H(4,6|4,4)$ is the (4,4)-analytic matrix parameter,
and the parameter $\tau_r$ \lb{A14b} does not depend on harmonics.
The matrix $e^v$ realizes the harmonic transform of the gauge superfields
$ A^k_\pm, \bar{A}_{k\pm}$ in the central basis (CB) to the equivalent
set of harmonic gauge superfields in the analytic basis (AB).
We have partially fixed  the CB-gauge invariance in \p{D5}, and the
$\lambda$-gauge transformations of $v$ will be used below.

The dynamical SYM-equations in the bridge representation \p{D4}
are reduced to the following harmonic differential conditions for
the basic Grassmann connections:
\be
(D^1_2,\;D^2_3,\;D^1_3)\left(\cA^1_-(v),\;\bcA_{3-}(v)\right)=0\;.
\lb{Hanal}
\ee

Using the off-shell $(4,4)$-analytic $\lambda$-transformations one can
choose the non-supersym\-me\-tric nilpotent  gauge condition for the
superfield bridge
\bea
&&v=\tmo b^1+ \btmh \bar{b}_3 +\tmo\btmh d^1_3\;,\q v^2=\tmo\btmh
[\bar{b}_3,b^1]\;,\q v^3=0~,\lb{D12c}\\
&&e^{-v}= I - v+{1\over2}v^2=I-\tmo b^1- \btmh \bar{b}_3+\tmo\btmh
({1\over2}[\bar{b}_3,b^1]-d^1_3 ) \;,
\lb{D12b}
\eea
where the fermionic   matrices $b^1, \bar{b}_3 $ and the bosonic matrix
$d^1_3 $ are analytic functions of the coordinates $\zeta$.

Note that the nilpotent gauges for the harmonic bridges and connections
are possible for the harmonic formalisms with the off-shell analytic
gauge groups only [3,7].

The  conditions for the $SU(n)$-bridge are
\be
\T v=0\;,\qq v^\dagger=-v\;.\lb{D15}
\ee

The matrices  $b^1, \bar{b}_3 $ and $d^1_3 $ have
the following properties in the gauge group $SU(n)$:
\bea
&& \T b^1=0\;,\q \T \bar{b}_3=0\;,\q \T d^1_3=0\;
,\lb{D13}\\
&&(b^1)^\dagger=\bar{b}_3\;,\q (d^1_3)^\dagger=-d^1_3
\;.\lb{D14}
\eea

Let us parameterize the Grassmann connection $\cA^1_-(v)$ and
$\bcA_{3-}$ in terms of the basic analytic matrices $b^1, \bar{b}_3$ and 
$d^1_3$ \p{D12c}
\bea
&&\cA^1_-(v)\equiv e^{-v}\Dmo e^v=b^1-\tmo (b^1)^2+\btmh f^1_3
+\tmo\btmh[b^1,f^1_3]\,, \lb{K9}\\
&&\bcA_{3-}\equiv e^{-v}\bDmh e^v=\bar{b}_3 +\tmo \bar{f}^1_3-
\btmh (\bar{b}_3)^2
+\tmo\btmh[\bar{f}^1_3,\bar{b}_3]~,
\eea
where the following auxiliary superfields are introduced:
\be
f^1_3=d^1_3-{1\over2}\{b^1,\bar{b}_3\}~,\q\bar{f}^1_3=-d^1_3-{1\over2}
\{b^1,\bar{b}_3\}~.\lb{K9b}
\ee

  Equations $(\Dot, \Dth) \cA^1_-(v)=0$ generates the following
independent relations for the (4,4)-analytic matrices:
\bea
&& \Dot b^1=-\tmt (b^1)^2\;,\lb{D17}\\
&&\Dth b^1=-\btmt f^1_3~,\lb{D18}\\
&&\Dot f^1_3=\tmt[f^1_3,b^1]~,\q \Dth f^1_3=0~.
\lb{D19}
\eea

  Equations $(\Dot, \Dth) \bcA_{3-}(v)=0$ are equivalent to the
relations
\bea
&&\Dot \bar{b}_3=\tmt\bar{f}^1_3~,\q \Dth \bar{b}_3=\btmt (\bar{b}_3)^2~,\\
&&\Dot\bar{f}^1_3=0~,\q \Dth\bar{f}^1_3=\btmt[\bar{b}_3,\bar{f}^1_3]~.
\eea
In the case of   gauge group $SU(n)$, the last  equations are not
independent, they can be obtained by conjugation from the equations
for $b^1$ and $f^1_3$.

It is useful to derive the following relations for the matrices $b^1$
and $\bar{b}_3$ which do not contain the auxiliary matrices $d^1_3, f^1_3$ or
$\bar{f}^1_3$:
\bea
&&\tmt\Dot(b^1,~\bar{b}_3)=0~,\q \btmt\Dth ( b^1,~\bar{b}_3)=0~,
\lb{D23}\\
&&\tmt\Dth b^1+\btmt\Dot \bar{b}_3=\tmt\btmt\{b^1,\bar{b}_3\}~.\lb{D24}
\eea

Solutions of the linear equations for  matrices $f^1_3$ and
$\bar{f}^1_3$ satisfy the subsidiary condition
\be
f^1_3 +\bar{f}^1_3=-\{b^1,\bar{b}_3\}~.
\ee

It is important that all these equations contain the nilpotent
elements $\tmt$ or $\btmt$ in the nonlinear parts, so they can be
reduced to the set of linear iterative equations using the
partial Grassmann decomposition.
In particular, the nilpotency of the basic equations yelds
 the subsidiary linear conditions  for the coefficient functions
\be
\Dot\Dot (b^1, \bar{b}_3,  d^1_3)=0\;,\q \Dth\Dth (b^1, \bar{b}_3,  d^1_3)
=0\;.\lb{lin2}
\ee

The  harmonic linear equations for the analytic superfields
$b^1, \bar{b}_3,  d^1_3$ have simple (short) solutions with the
finite number of the harmonic on-shell field components
\footnote{  The analogous short harmonic $N=3, 4$ Abelian superfields
have been considered in Ref.[13].}.
This harmonic shortness is an important restriction on the structure
of  the SYM-solutions.

\section{\lb{H}Analytic representation of solutions }

Remember that the following covariant Grassmann derivatives are flat in
the analytic representation of the gauge group
before the gauge fixing:
\be
e^v\nabla^1_\pm e^{-v}\equiv\hat\nabla^1_\pm=D^1_\pm\;,\q 
e^v\bar{\nabla}_{3\pm} e^{-v}\equiv\hat{\bar{\nabla}}_{3\pm}
=\bar{D}_{3\pm}\;\lb{H0}
\ee

The harmonic transform of the covariant derivatives using the matrix
$e^v$ \p{D4} determines the composed on-shell harmonic AB-connections
\bea
&&\nabla^I_K\equiv e^v D^I_K e^{-v}=D^I_K+V^I_K(v)~,\nn\\
&&V^I_K(v)=e^v(D^I_K e^{-v})\lb{D10}~.
\eea
Note that the harmonic connections in the bridge representations satisfy
automatically the harmonic
zero-curvature equations.

It is evident that the basic equations \p{Hanal} are equivalent
to the following set of the dynamic G-analyticity relations
for the composed harmonic connections:
\be
 (\Dmo,~\bDmh) \left(V^1_2(v), V^2_3(v), V^1_3(v)\right)=0\;,\lb{D7}
\ee

The positive-helicity analyticity conditions
are satisfied automatically for the bridge in the gauge  \p{D5}.

The dynamical G-analyticity equation \p{D7} in  gauge \p{D12c}
is equivalent to the following harmonic differential bridge equation:
\bea
&&V^1_2(v)=\tmt b^1\;,\qq (V^1_2)^2=0\;\lb{dyn}\\
&&\Dot e^{-v}=e^{-v}V^1_2~.\lb{br}
\eea
where the manifestly analytic nilpotent representation for the harmonic
connection is used.

The 2-nd on-shell harmonic connection is also nilpotent
$V^2_3(v)=-\btmt \bar{b}_3$.

The spinor AB-connections can be calculated via the
non-analytic harmonic connections  by analogy with
the $N=2$ formalism of Ref.[8]
\bea
&&a^2_\pm=-D^1_\pm V^2_1~,\q a^3_\pm=-D^1_\pm V^3_1~,\\
&&\bar{a}_{2\pm}=\bar{D}_{3\pm} V^3_2
\;,\q \bar{a}_{1\pm}=\bar{D}_{3\pm} V^3_1\lb{H0b}
\eea
where $V^2_1, V^3_1$ and $V^3_2$ are the non-analytic harmonic connections.

 The  3-rd harmonic analytic connection can be readily calculated
\be
V^1_3=\Dot V^2_3-\Dth V^1_2+[V^1_2,V^2_3]=\tmh b^1-\btmo \bar{b}_3
~,\lb{3con}
\ee
where Eq.\p{D24} is used.

  The harmonic connection $V^2_1$ can be
written in terms of the superfield $b^1$  only
\be
V^2_1=-\tmo \Dto b^1~.\lb{nonan4}
\ee

The conjugated connection
$V^3_2=-(V_1^2)^\dagger$ contains  matrix $\bar{b}_3$ only.

These connections  satisfy  the partial G-analyticity conditions
\be
\bar{D}_{3\pm}V^2_1=0\;,\q D^1_\pm V^3_2=0 \lb{H2}~.
\ee

It is convenient to define the  AB-superfield strengthes
\be
\bar{w}^{12}=-\Dpo\Dmo V^2_1
=-\Dpt b^1\;,
\lb{H5}
\ee
where Eq.\p{nonan4} is used.

Stress that the single coefficient matrix $b^1$ generates the family
of the AB-geometric objects: $V^1_2, V^2_1, a^2_\pm$ and $\bar{w}^{12}$.
The conjugated $\bar{b}_3$-family of superfields contains $V^2_3, V^3_2,
\bar{a}_{2\pm}$ and $w_{23}$.

The superfield $\bar{w}^{12}$ satisfy the (4,2)-dimensional G-analyticity
conditions
\be
D^1_\pm\bar{w}^{12}=\bar{D}_{3\pm}\bar{w}^{12}=D^2_\pm\bar{w}^{12}+
[a^2_\pm,\bar{w}^{12}]=0
\ee
and the non-Abelian H-analyticity conditions
\be
(\nabla^1_2, \nabla^2_3, \nabla^1_3)\bar{w}^{12}=0~.
\ee
The (4,2)-analytic superspaces have been considered earlier in Refs.[12,13].

It should be noted that  function $d^1_3$ (or $f^1_3$) is an auxiliary
quantity in the framework of the analytic basis, since all harmonic
and spinor connections and tensors of this basis can be expressed in terms
of  superfields $b^1$ and $\bar{b}_3$ only. Nevertheless, the construction
of $d^1_3$ is important for the transition to the central basis.

Let us analyze the   analytic equations  for the basic fermionic
(4,4) matrices $b^1$ and $\bar{b}_3$. The nonlinear terms in these equations 
contain the negative-helicity Grassmann coordinates
$\tmt, \tmh, \btmo$ and $\btmt$, so the partial decomposition in terms of
these coordinates is very useful for the iterative analysis of solutions.
Equations for the auxiliary matrix $d^1_3$ do not give additional
restrictions on $b^1$ and $\bar{b}_3$.

 Consider first the decomposition of the  harmonic derivatives and define
the harmonic derivatives  on the (4,0) analytic functions depending on
the analytic Grassmann variables  $\tpt,\tph,\btpo,\btpt$
\bea
&&\hDot=\partial^1_2+i\tpt\btpo\Pp-\tpt\ppo+\btpo\bppt~,\nn\\
&&\hDth=\partial^2_3+i\tph\btpt\Pp-\tph\ppt+\btpt\bpph~,\nn\\
&& \hDoh=\partial^1_3+2i\tph\btpo\Pp
-\tph\ppo+\btpo\bpph~.
\lb{I1}
\eea

The (4,0) decomposition of the  (4,4) analytic matrix functions has the
following form:
\bea
&&b^1=\beta^1+\tmt B^{12}+\tmh B^{13}+\btmo B^0+\btmt B^1_2
+\tmt\tmh \beta^0+ \tmt\btmo \beta^2+\tmh\btmt \beta^{13}_2\nn\\
&&+\tmh\btmo \beta^3+\btmo\btmt \beta_2
+\tmt\btmt\eta^1+\tmt\tmh\btmo B^{23}
+\tmh\btmo\btmt  B_2^3+\tmt\tmh\btmt C^{13}\nn\\
&&+\tmt\btmo\btmt C^0+\tmt\tmh\btmo\btmt\eta^3~.\lb{I2}
\eea
The analogous decompositions can be written for $\bar{b}_3$ and $d^1_3$.

It is easily to show that a part of the (4,0) coefficients
can be constructed as the algebraic functions of the basic set of
independent (4,0) matrices, for instance,
\bea
&& B^1_2=-\hDot B^0-i\tpt\PY\beta^1~,\lb{I4}\\
&&B^{12}=\hDth B^{13}+i\btpt\bPY\beta^1~,\lb{I5}\\
&&\beta^2=\hDth \beta^3-i\btpt\bPY B^0~,\lb{I6}\\
&&\eta^1=-\hDot\beta^2+i\btpo\bPY B^0
+i\tpt\PY B^{12}
+[\beta^1,B^0]
~.\lb{I8}
\eea

The independent (4,0) matrix functions are
\be
B^0, B^{13}, B^{23}, (C^0-\bar{C}^0), B^3_2, \beta^1, \beta_2, \beta^3,
\beta^0,\eta^3~.\lb{I12}
\ee

The (4,0) matrix of the dimension $l=-1/2$ satisfies the linear
equations
\be
\hDot\beta^1= \hDth \beta^1=0~.\lb{I13}
\ee

The equations for the $l=-1$ matrices $B^{13}$ and $B^0$ are
\bea
&& \hDot B^{13}=0~,\q(\hDth)^2 B^{13}=0~,\nn\\
&&\hDoh B^{13}=-2i\btpo\bPY\beta^1-(\beta^1)^2~,\lb{I15}\\
&&\hDth B^0=0~,\q(\hDot)^2 B^0=0~,\nn\\
&&\hDoh (B^0-\bar{B}^0)
=i\tph\PY\beta^1+i\btpo\bPY\bar\beta_3+
\{\beta^1,\bar\beta_3\} ~.\lb{I16}
\eea
The inhomogeneous linear equations for $B^{13}$ and $B^0$ contain
sources with the functions $\beta^1$ and $\bar{\beta}_3=(\beta^1)^\dagger$
calculated on the previous stage.

The iterative equations for the (4,0) matrices with $l < -1$ can be
analyzed analogously.
Each independent iterative equation is manifestly resolved in terms of
the harmonic derivatives of the corresponding  function and the
sources of these equations  can  be calculated on the previous stage of 
iteration.

Thus, it can be shown easily that the basic  $N=3$ harmonic  equations
for the (4,4) analytic moduli functions  with the nilpotent nonlinear
terms are equivalent to the finite number of the
linear iterative (4,0) equations which contain non-Abelian sources
constructed from the solutions of the previous step of iteration.

Note that all iterative equations are simplified essentially for
the two-dimensional solutions which do not depend on  variables
$Y$ and $\bar{Y}$.

\section{Conclusions and acknowledgment}

We have described the harmonic transform of the $N=3$ SYM-equations of motion
in the standard superspace to the  differential HSS equations
for the bridge matrix $e^v$ which connects different representations
of the gauge superfields. The light-cone nilpotent 
 gauge condition for  matrix $v$ simplifies significantly
the analysis of the bridge equations. This condition yields the
nilpotent harmonic-analytic gauge connections and the
corresponding linear 2-nd order differential conditions for the basic
matrices. The nilpotency of  nonlinear terms in the basic HSS equations
allow us to consider the simple iterative procedure based on the partial
decomposition in Grassmann variables of the negative helicity.
The finite set of these  solvable linear iterative equations can be used
for the explicit construction of the $N=3$ SYM-solutions in the harmonic
superspace and the on-shell gauge superfields in the ordinary superspace.

Let us discuss shortly the problems of solving  equations of the
$N=0, 1$ and 2 subsystems of the $N=3$ theory.  One can study the problem 
of constructing solutions with the reduced number of fermion or scalar 
fields using the explicit constructions of the $N=3$ solutions, although  
the formal reduction of the stable and regular $N=3$ solution could, 
in principle, correspond to unstable or irregular solutions with lower 
supersymmetry.

 The $N=2$  SYM-constraints together with equations of motion have the
 Grassmann-harmonic zero-curvature representation in the framework of 
the $SU(2)$-harmonic superspace [5]. It should be underline
that the nilpotent bridge representation similar to \p{D12c} is also very 
useful for solving the $N=2$  SYM-equations. The corresponding $N=2$ 
solutions will be discussed elsewhere.

The author is grateful to J. Niederle for  collaboration and to 
E. Ivanov and E. Sokatchev for  discussions.

This work is supported by the Votruba-Blokhintsev programme
 in Joint Institute for Nuclear Research and also by
the grants RFBR-99-02-18417, RFBR-DFG-99-02-04022
and NATO-PST.CLG-974874.

\newcommand{\etal}{{\em et al.}}
\setlength{\parindent}{0mm}
\vspace{5mm}
{\bf References}
\begin{list}{}{\setlength{\topsep}{0mm}\setlength{\itemsep}{0mm}%
\setlength{\parsep}{0mm}}
%
\item[1.]  M.F. Sohnius,  Nucl. Phys. {\bf B 136},  461 (1978).
\item[2.]  E. Witten, Phys. Lett. {\bf B 77}, 394 (1978). 
\item[3.] A. Galperin, E. Ivanov, S. Kalitzin, V. Ogievetsky and
E. Sokatchev,  Quant. Grav. {\bf 2}, 155 (1985). 
\item[4.] B.M. Zupnik,
Sov. J. Nucl. Phys. {\bf 48}, 744 (1988); Phys. Lett. {\bf B 209}, 513 (1988). 
\item[5.] B.M. Zupnik, Phys. Lett. {\bf B 375}, 170 (1996).
\item[6.]  C. Devchand and V. Ogievetsky, Integrability of N=3
super-Yang-Mills equations, hep-th/9310071.
\item[7.] A. Galperin, E. Ivanov, S. Kalitzin, V. Ogievetsky and
E. Sokatchev, Class. Quant. Grav. {\bf 1}, 469 (1984).
\item[8.] B.M. Zupnik,  Phys. Lett. {\bf B 183}, 175 (1987).
\item[9.] C. Devchand and A.N. Leznov, Comm. Math. Phys.
{\bf 160}, 551 (1994).
\item[10.] J.-L. Gervais and M.V. Saveliev,  Nucl. Phys. {\bf B 554}, 183
 (1999).
\item[11.] J. Niederle and B. Zupnik, Harmonic-superspace transform
 for N=3 SYM-equations, hep-th/0008148.
\item[12.]  A. Galperin, E. Ivanov and V. Ogievetsky, Sov. J. Nucl.
Phys. {\bf 46}, 543 (1987). 
\item[13.] L. Andrianopoli, S. Ferrara, E. Sokatchev, B. Zupnik,
Shortening of primary operators in N-extended $SCFT_4$ and harmonic-
superspace analyticity, hep-th/9912007.
%
\end{list}
\end{document}